\documentclass[a4paper,twoside,10pt]{article}
\usepackage{graphicx,publaob}

\def\papertype{Invited Review}
\begin{document}

% Title
\title{MONITORING BLAZAR VARIABILITY TO UNDERSTAND EXTRAGALACTIC JETS}

% Authors
\authors{CLAUDIA MARIA RAITERI$^1$}

% Addresses and e-mails
\address{$^1$INAF-Osservatorio Astrofisico di Torino, \break via Osservatorio 20, I-10025 Pino Torinese, Italy}
\Email{claudia.raiteri}{inaf}{it}

% Running titles
\markboth{\runningfont MONITORING BLAZAR VARIABILITY TO UNDERSTAND EXTRAGALACTIC JETS}
{\runningfont C. M. RAITERI}

\vskip-1mm

\abstract{We review the main topics in the field of blazar multiwavelength variability as a tool to understand the physics and structure of extragalactic jets and their central engine. We address issues such as the cross-correlation between flux variations at different frequencies, the mechanisms to explain the long-term and short-term variability, the size of the jet emitting regions, the polarization behaviour, and the periodicities detected in the multiwavelength light curves.}

\vskip-.5cm

\section{INTRODUCTION}

Blazars are a peculiar class of jetted active galactic nuclei, with one relativistic jet pointing towards us. As a consequence, the jet radiation undergoes relativistic beaming, whose effects can be expressed in terms of the Doppler factor $\delta=[\Gamma(1-\beta \cos\theta)]^{-1}$, where $\Gamma=1/\sqrt{1-\beta^2}$ is the Lorentz factor, $\beta$ is the plasma velocity normalized to the speed of light, and $\theta$ is the viewing angle. The Doppler factor thus depends on two quantities: the plasma speed and the orientation of the emitting region with respect to the line of sight. The main effects of the Doppler beaming are that the flux that we observe is enhanced in comparison to what is emitted by the source, the frequencies are blue-shifted, and the variability time scales are shortened (e.g.~Urry \& Padovani 1995).

Blazar emission is observed over the entire electromagnetic spectrum, from the radio band to the most energetic $\gamma$ rays. 
Because of the Doppler beaming, the broad-band spectral energy distribution (SED) of blazars is dominated by the non-thermal radiation from the jet.
%, which usually overcomes the other contributions.
In the usual $\log \nu F_\nu$ versus $\log \nu$ diagram, the SED of the jet shows two bumps.
The low-energy bump is due to synchrotron radiation produced by electrons spiralling around the magnetic field lines in the jet. 
The nature of the high-energy bump of the SED is more debated (e.g.~de Jaeger et al. 2023). The most popular explanation involves inverse-Compton scattering of low-energy photons off the same relativistic electrons that produce the synchrotron radiation. However, the fact that blazar jets are possible emitters of high-energy neutrinos (e.g.~Giommi et al. 2020) suggests that the high-energy emission receives contributions also from other processes, involving hadrons instead of leptons.

The class of blazars includes two types of sources that differ according to the strength of their emission lines. The spectra of BL Lacertae-type (BL Lac) objects show only weak lines or no lines at all, whereas the spectra of flat-spectrum radio quasars (FSRQs) usually exhibit strong lines.
Indeed, in addition to the nonthermal jet emission, which is very variable and polarized, the radiation from FSRQs can receive an important emission contribution from the so-called ``big blue bump", which is due to the accretion disc and broad and narrow line regions, and which is less variable and not polarized. The big blue bump peaks in the UV in the rest frame and, depending on the redshift, can affect the optical and even the infrared spectral regions.
In close BL Lacs, we can instead have a strong nonvariable and nonpolarized contribution from the stellar light in the host galaxy, which shows up in the optical--infrared region.\\

\vskip-3mm

\section{CROSS-CORRELATIONS BETWEEN BANDS}

Blazar light curves are characterized by unpredictable variability at all wavelengths and on all time scales, from years to minutes; in general, fast flux changes overlap long-term oscillations.
The comparison between the source behaviour at different frequencies is a very useful tool to derive a wealth of information on the emission and variability mechanisms and location of the emitting regions in the jet. 

\begin{figure}
	\centering
\includegraphics[width=7.3cm,keepaspectratio=true]{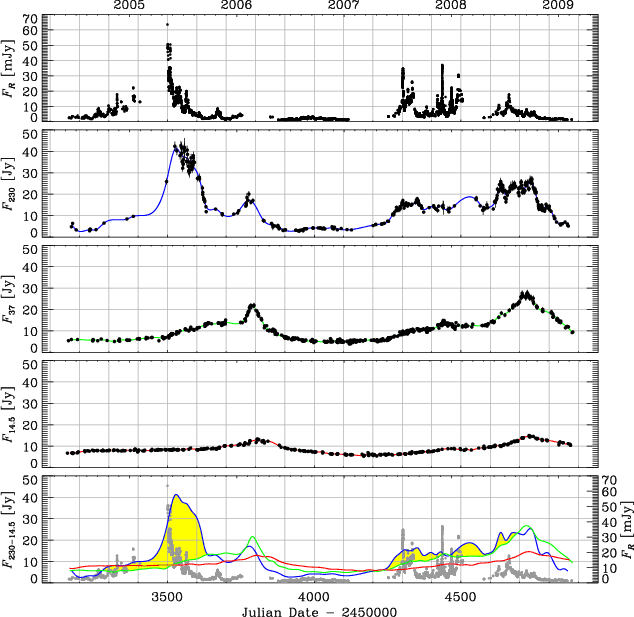}
\includegraphics[width=5.5cm,keepaspectratio=true]{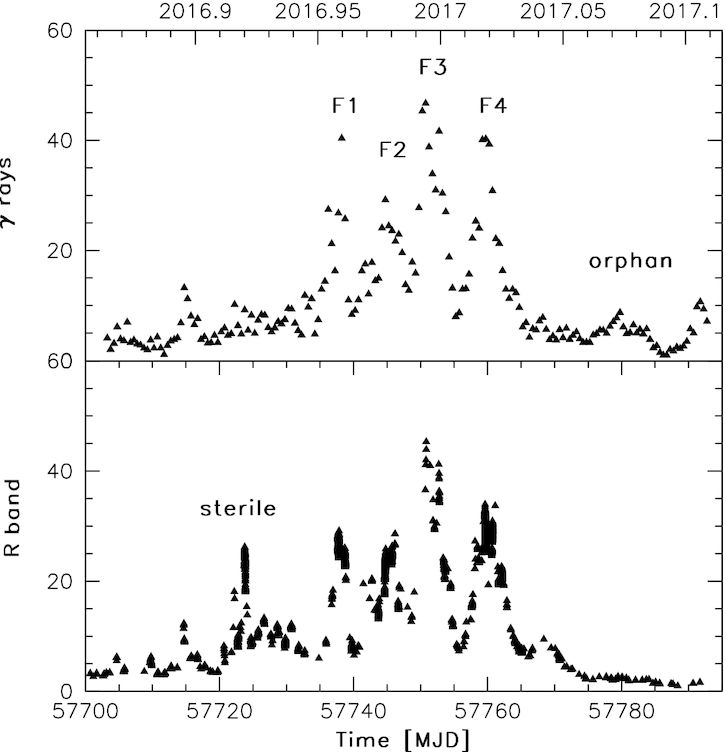}

\vspace{-3mm}

\caption{\emph{Left}: From top to bottom, flux densities of 3C~454.3 in the $R$ band (i), at 230 GHz (ii), 37 GHz (iii), and 14.5 GHz (iv), and comparison between the trends of the above light curves (v). From Villata et al. (2009). \emph{Right}: {\it Fermi} $\gamma$-ray (top) and optical (bottom) light curves of CTA~102 highlighting an ``orphan" $\gamma$-ray flare and a ``sterile" optical flare. From D'Ammando et al. (2019).}
	\label{cross}
\vspace{-1mm}
\end{figure}

In the case of the optical and radio bands (see Figure~\ref{cross}), correlations are often found, and in these cases the radio flux variations follow the optical ones, with a delay that increases with wavelength.
This most likely means that the jet is inhomogeneous, and that radiation with increasingly longer wavelength is emitted from more and more external jet regions (e.g. Villata et al. 2009).
Moreover, the flux variations at longer wavelengths are smoother, and this suggests that larger jet regions are involved (e.g. Raiteri et al. 2017).

In the case of $\gamma$ and optical fluxes (see Figure~\ref{cross}), they are usually observed to vary roughly simultaneously or with a small delay, as expected if the $\gamma$ rays are produced by inverse Compton scattering of soft photons off the same electrons that are responsible for the optical radiation. 
But there are exceptions to this behaviour. Sometimes ``orphan” $\gamma$-ray flares are detected, without an optical counterpart. And sometimes we observe ``sterile" optical flares, which do not give rise to a $\gamma$-ray flare. In these cases, we may think that there are different emitting regions in the jet, or that the $\gamma$-ray event was produced by a hadronic process (e.g. D'Ammando et al. 2019, de Jaeger et al. 2023).\\

\vskip-3mm

\section{THE WHOLE EARTH BLAZAR TELESCOPE}

To obtain robust results on flux correlations and their time delays, we need continuous, well-sampled light curves. In the optical band, this can be reached only thanks to the common effort of many observers, joining wide international collaborations such as the Whole Earth Blazar Telescope\footnote{https://www.oato.inaf.it/blazars/webt} (WEBT). 

The WEBT was born in the Nineties, when the {\it Compton Gamma Ray Observatory} (CGRO) satellite discovered that the extragalactic $\gamma$-ray sky is full of blazars.
In 1997 John Mattox (Boston University) had the idea to bring together optical observers from all over the world to provide ground-based support to the space $\gamma$-ray observations of CGRO. The aim was to build a network with good coverage in longitude to shift the observing task from East to West as the Earth rotates and get ideally uninterrupted monitoring. 
In 2000 the WEBT coordination moved to Massimo Villata (Osservatorio Astronomico di Torino\footnote{Now INAF - Osservatorio Astrofisico di Torino}) as President and to me as Executive Officer.
Radio and near-IR astronomers were invited to join the WEBT to cover a wider range of wavelengths.

In 26 years of activity, many broad-band multiwavelength campaigns have been organized to investigate the blazar behaviour at low and high energies.
About 200 observers participated in the WEBT campaigns, acquiring data with more than 150 telescopes, mostly in the optical band. 
%There are WEBT members in Africa, America, and Asia, but most of them are in Europe, including the Serbian colleagues from the Astronomical Observatory of Belgrade.
Some high-level nonprofessional astronomers also contribute to the observations of bright objects. 
Moreover, we had and have very fruitful collaborations with a number of teams involved in high-energy observations, especially with the {\it AGILE}, {\it Fermi}, and MAGIC teams. 
{\it AGILE} and {\it Fermi} are satellites that observe the sky in the energy ranges 18--60 keV plus 30 MeV--50 GeV (AGILE), and 10 keV--300 GeV (Fermi). MAGIC is a stereo Cherenkov telescope sensitive to photon energies between 30 GeV to 100 TeV.

The WEBT can collect a huge amount of data, providing mostly photometry, but also polarimetry, and in some cases spectroscopy.
To complement the low-energy data acquired during the WEBT campaigns, we sometimes obtained coordinated observations from space, in particular with the {\it XMM-Newton}, {\it Swift}, and {\it TESS} satellites.
The WEBT data are stored in an archive and made available to interested researchers after publication. 
%In particular, the WEBT archive contains more than 2200 data from the Vidojevica telescopes starting from 2012.
In our papers, we are also proposing models to explain blazar variability.
%There are 267 papers in the NASA ADS, either led by the WEBT, or to which the WEBT contributed data. About half of these papers are refereed, including 3 papers on Nature, 2 of which led by the WEBT Collaboration. 
Many of the results mentioned in this proceeding were obtained by the WEBT Collaboration.\\

\vskip-3mm

\section{VARIABILITY MECHANISMS}

When we see flares in blazar light curves, we can imagine that particles get accelerated in the jet. The two main acceleration mechanisms that are currently much debated are: shock waves propagating in the jet (e.g.~Marscher \& Gear 1985), and magnetic reconnection (e.g.~Sironi \& Spitkovsky 2014), possibly triggered by kink instabilities. 

Moreover, as mentioned in the Introduction, Doppler beaming depends on the viewing angle, so strong flux variations are expected if the jet-emitting regions change their orientation with respect to the line of sight. This can happen because of jet precession, or rotation induced by orbital motion in a supermassive black hole (SMBH) binary system, or jet twisting due to kink instabilities developing inside the jet. 
There are many examples of both observational (e.g.~Lico et al. 2020, Zhao et al. 2022) and theoretical (e.g.~Nakamura et al. 2001, Moll et al. 2008, Acharya et al. 2021) results indicating that the jet can indeed be curved and twisting.

\begin{figure}
	\centering
\includegraphics[width=6.2cm,keepaspectratio=true]{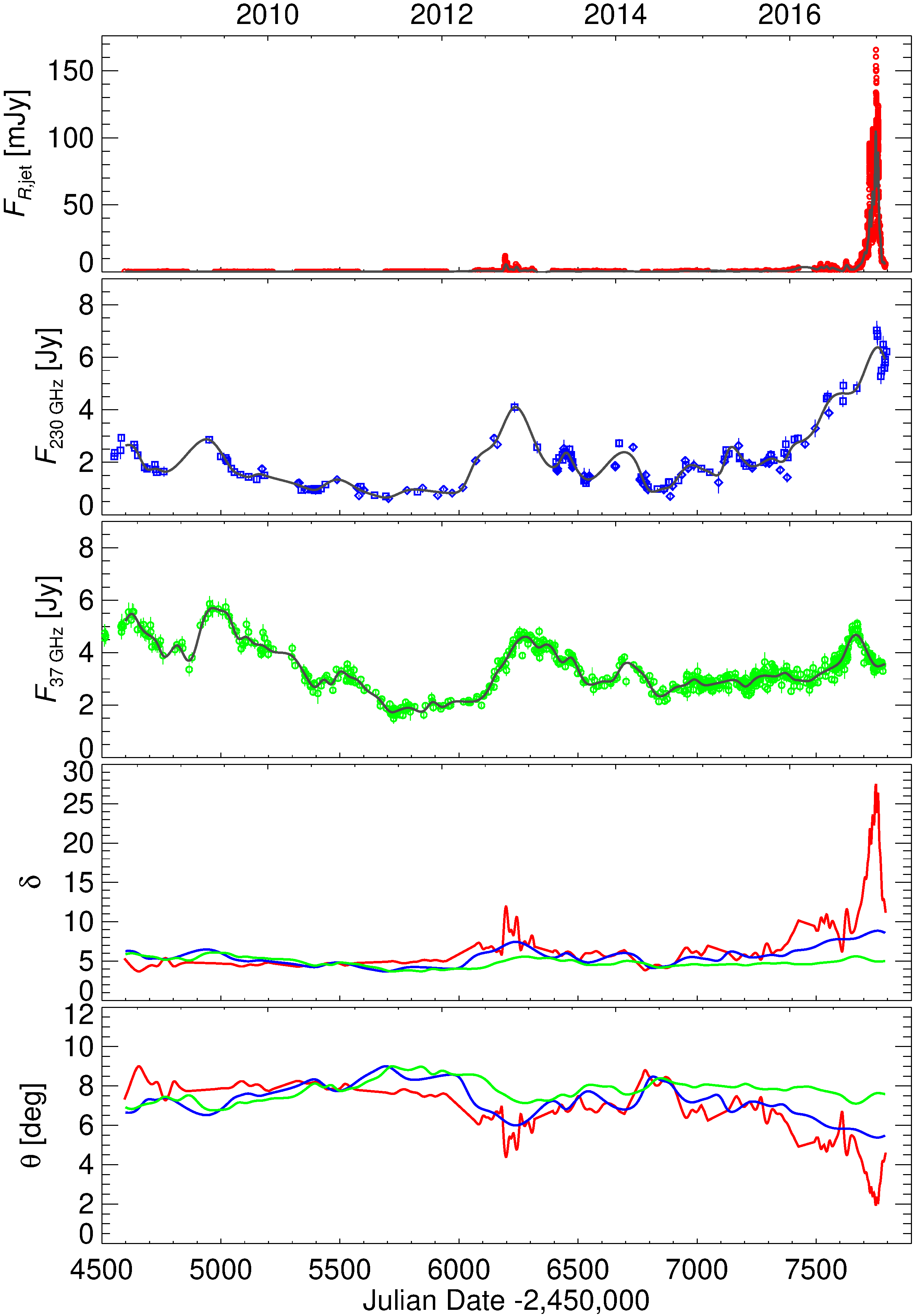}
\includegraphics[width=6.6cm,keepaspectratio=true]{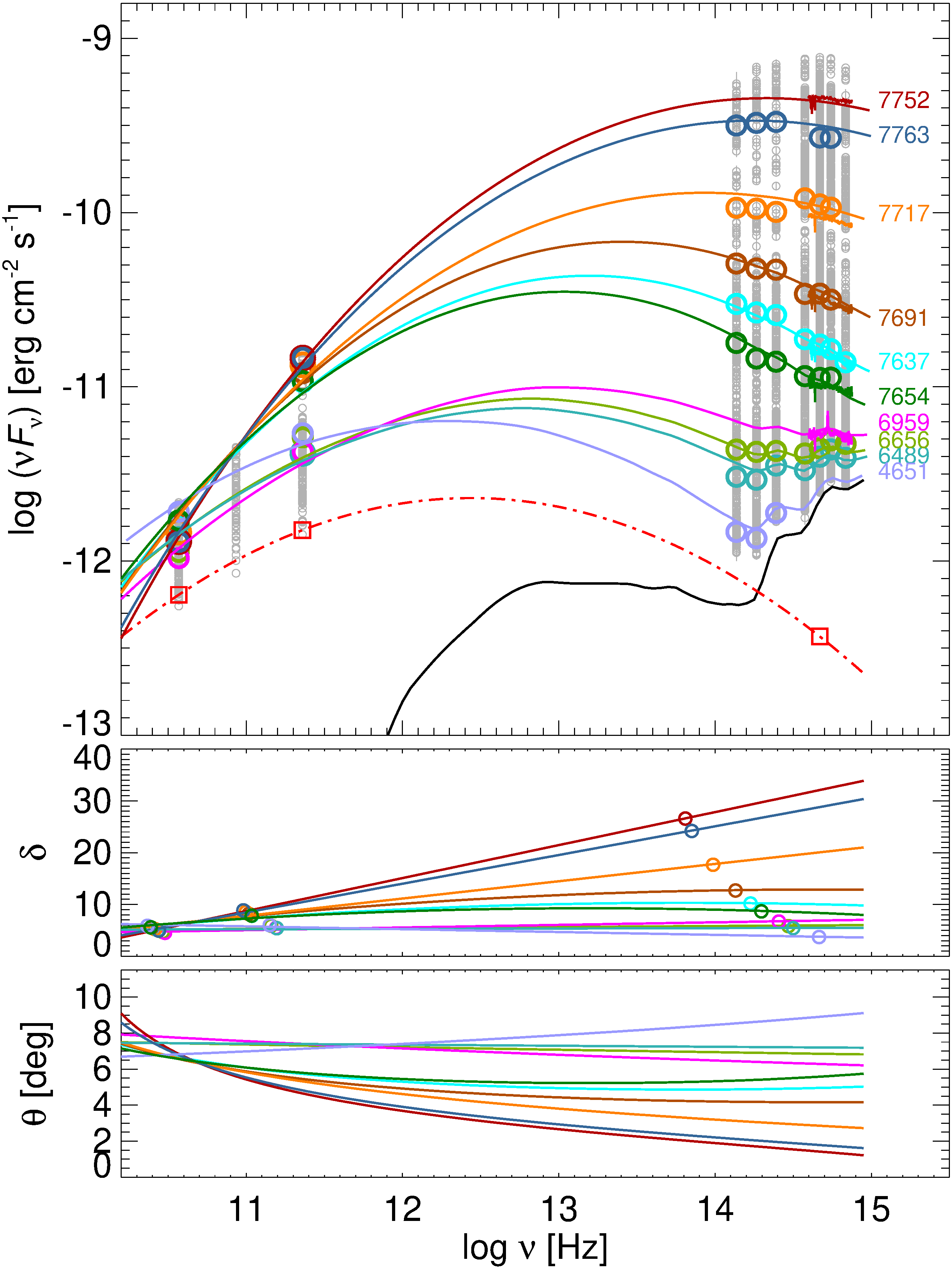}

\vspace{-3mm}

\caption{\emph{Left}: From top to bottom, flux densities from the CTA 102 jet in the $R$ band (i), at 230 GHz (ii), and at 37 GHz (iii), and behaviour in time of the Doppler factor $\delta$ (iv) and of the viewing angle $\theta$ (v) for the corresponding jet emitting regions. \emph{Right}: From top to bottom, spectral energy distributions for CTA~102 in different brightness states (i); empty circles represent observations, continuous lines of the same colour show predictions of the twisting jet model. Behaviour along the jet of $\delta$ (ii) and $\theta$ (iii); synchrotron frequencies are assumed to increase towards the black hole. Adapted from Raiteri et al.~(2017).}
	\label{cta102}
\vspace{-1mm}
\end{figure}

In several papers by the WEBT Collaboration we have proposed that the long-term trend in blazar multiwavelength light curves is due to changes in the jet orientation, which produce changes in the Doppler factor and hence in the flux. This geometrical model was detailed in Raiteri et al. (2017), where we analysed the optical and radio behaviour of the FSRQ CTA~102 over more than 8 years, including the extraordinary outburst observed at the end of 2016 -- beginning of 2017.
We proposed a jet which is inhomogeneous, which means that radiation at different frequencies is emitted from different regions, the longer the wavelength, the more external its production site in the jet. This jet is also curved, which means that different emitting regions have different viewing angles, and thus the radiation in the various zones undergoes different Doppler beaming. Finally, the jet is twisting, which means that the viewing angle varies in time because of internal instabilities and/or external reasons, such as orbital motion or precession. 

If we then assume that the jet is inhomogeneous, curved, and twisting, from the light curves at different frequencies we can derive the behaviour in time of the Doppler factor and then the behaviour in time of the viewing angle for the emitting regions in the jet, as shown in Figure~\ref{cta102}. This means that we can reconstruct the twisting motion along the jet. And this model can also reproduce the shape of the SED at various brightness levels, allowing us to also infer the trend of the Doppler factor and viewing angle as a function of the emitted frequency (see Figure~\ref{cta102}), i.e., moving along the inhomogeneous emitting jet , from the inner emitting regions, where more energetic synchrotron photons are produced, to the outer zones, where less energetic photons are generated.\\

\vskip-3mm

\section{INTRADAY VARIABILITY}

The geometrical interpretation can be applied successfully to the long-term variability.
But blazars are known to also show IntraDayVariability (IDV), i.e.~variability on sub-daily time scales, and even microvariability, i.e.~variability on sub-hour time scales.
IDV has actually been known for about 50 years and has extensively been studied at optical and radio wavelengths (see Wagner \& Witzel 1995 for a review).

From causality arguments based on the light travel time, the shortest observed time scale $\Delta t_{\rm obs}$ can set an upper limit to the dimension of the emitting regions: $R < c \Delta t_{\rm obs} \delta/(1+z)$, where $c$ is the light speed and $z$ is the source redshift.
The jet cannot be too compact, otherwise the photons would get absorbed. 
The size of the jet at the point where dissipation occurs is typically estimated as $\sim 10^{16}$ cm (e.g.~Ghisellini et al. 2007).
Microvariability then implies that either the Doppler factor must be very high, sometimes higher than 50 or even 100, which seems to be in contradiction with the observations, or the emission that we observe must come from jet substructures.

\begin{figure}
	\centering
\includegraphics[width=6.4cm,keepaspectratio=true]{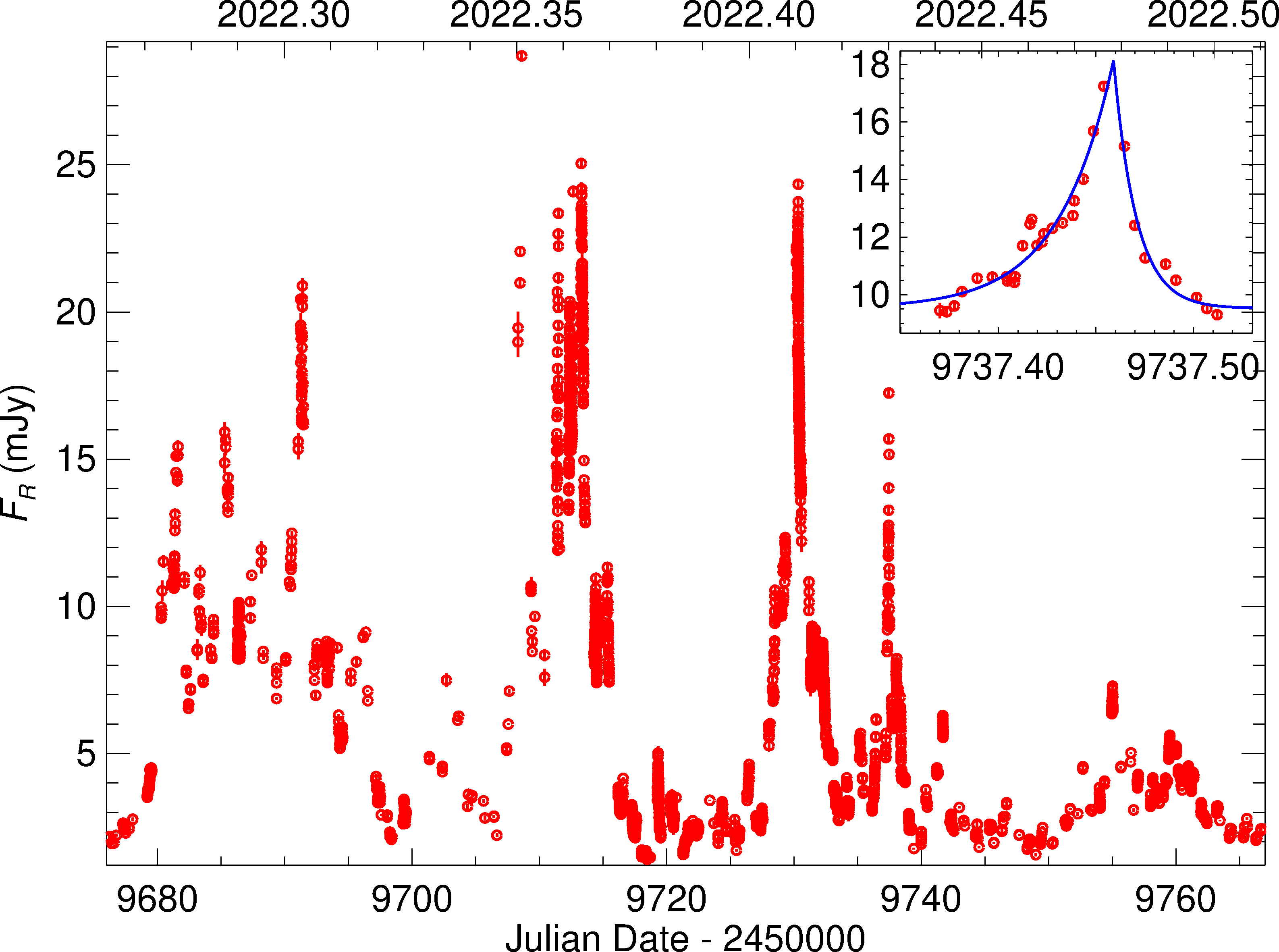}
\includegraphics[width=6.4cm,keepaspectratio=true]{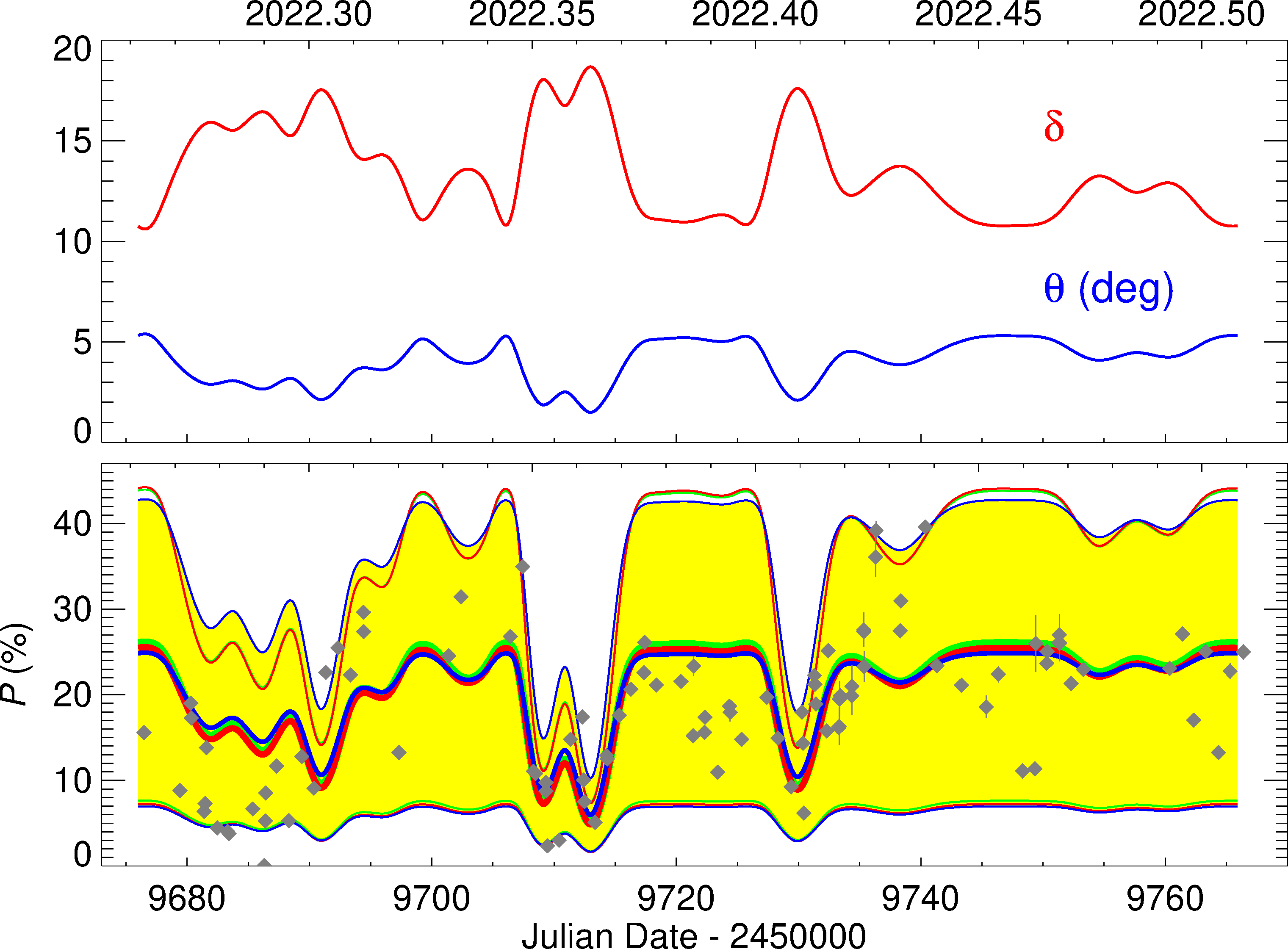}

\vspace{-3mm}

\caption{\emph{Left}: $R$-band light curve of the BL Lac object S4~0954+65 obtained by the WEBT during the strong activity period in 2022. The inset shows the results of modelling one of the fastest flares. \emph{Right}: The behaviour in time of $\delta$ and $\theta$ (top) and of the polarization degree $P$ (bottom) with the outcome of three different models overlapped. Adapted from Raiteri et al.~(2023).}
	\label{0954_flare}
\vspace{-1mm}
\end{figure}

Figure~\ref{0954_flare} shows the results obtained by the WEBT by following the BL Lac object S4~0954+65 during a period of extreme variability in 2022 (Raiteri et al. 2023). Modelling the fastest flare led to a variability time scale as short as 17 minutes, which implies an upper limit to the optical emitting region of the order of $10^{-4}$ parsec, i.e.~about one hundredth of the typical size of a blazar jet.

Recently, space missions like {\it Kepler} and {\it TESS}, with their exceptional continuous cadence, have confirmed the occurrence of microvariations in the optical band (see e.g.~Weaver et al. 2020, Raiteri et al. 2021a,b, Wehrle et al. 2023). 

Blazar microvariability with as short as a few minutes time scales was also detected at $\gamma$ rays, in both the GeV and TeV energy domains. In the case of 3C~279 observed by the {\it Fermi} satellite, a very fast flare in 2018 was ascribed to magnetic reconnection in a region of less than $10^{15}$ cm (Shukla \& Mannheim 2020). For two BL Lacs observed by HESS and MAGIC, the inferred dimensions are likely more than 10 times smaller (Aharonian et al. 2007, Albert et al. 2007).

The assumption that the jet emission is the result of the contributions from many sub-regions can be brought to its extreme consequences, leading to the concept of turbulence.
Turbulence can be modelled as a set of plasma cells  with largely independent properties. Marscher (2014) showed that the passage of shocks in turbulent plasma can reproduce the main features of blazar multiwavelength light curves and SEDs, and also the behaviour of polarization.\\

\vskip-3mm

\section{POLARIZATION}

Polarization adds an important piece of information to blazar variability studies, because polarization depends on the magnetic field, which determines the physics of extragalactic jets. 
Observations show that in blazars both the polarization degree (PD) and the polarization angle, namely the electric vector position angle (EVPA), are very variable.  To study their behaviour and correlation with flux, we must take into account that for some sources the jet flux may be increased, and the jet PD may be lowered, by the emission contributions from the big blue bump or from the host galaxy, whose radiation is not polarized. As for the EVPA, there is a 180 degree uncertainty on it, so we need a high observing cadence to be confident that we can reconstruct possible rotations of the angle in time in a reliable way. 

In any case, the interpretation of blazar polarimetric behaviour is very hard.
Sometimes we see that the PD is correlated with the flux, at least as a trend, sometimes it is anticorrelated, and sometimes we cannot find any correlation. 
Moreover, we often see wide rotations of the EVPA but, again, they show some correlation with the flux only in a few cases, so there is a debate on whether they are most likely of stochastic rather than deterministic nature.

Wide rotations of the EVPA have been observed for many years now.
Already in 1979 Ledden and Aller noticed a wide rotation of the radio EVPA in a BL Lac object during a radio flare, and suggested that this was due to a rotating magnetic field structure (Ledden \& Aller 1979).
A wide rotation of the polarization angle of 3C~279 was observed by Abdo et al. (2010)  simultaneously with a $\gamma$-ray flare.
In recent years, the Robopol project running at the Skinakas telescope in Greece has followed the polarimetric behaviour of a large sample of blazars to derive statistical properties. They claim that the rotations are mostly correlated with $\gamma$-ray flares, favouring a deterministic origin for them (Blinov et al. 2018).

However, there are a number of cases where wide EVPA rotations are observed in both directions, clockwise and anticlockwise, and do not appear to be correlated with the optical and $\gamma$-ray fluxes. 
For example, large rotations of the optical EVPA were observed in CTA~102 during faint states of the source, and not during its extraordinary outburst of 2016--2017, and also the PD during the outburst did not show any special behaviour (Raiteri et al. 2017). In these cases turbulence may have played an important role

In Raiteri et al. (2023), we tried to see whether we could explain the variability of the PD of S4~0954+65 in terms of the twisting jet model detailed in Raiteri et al. (2017). From the long-term trend of the flux we derived the variation in time of the Doppler factor and of the viewing angle of the optical emitting region, as shown in Figure~\ref{0954_flare}. We explored different models for the polarization behaviour, involving helical magnetic fields or shock waves. All these models basically predict that the PD depends on the square of the sin of the viewing angle in the rest frame, which is Lorentz-transformed in the observer's frame. In the bottom panel of Figure~\ref{0954_flare}, the thick lines highlight what we obtained in this way: we could reproduce the average trend, i.e. the general anticorrelation between flux and polarization degree. However, there is a large dispersion, so we must admit that other intrinsic processes are at work to order or disorder the magnetic field, such as turbulence, shocks, or magnetic reconnection. These intrinsic, energetic processes are in any case required to explain the fast changes of the flux, since the geometrical model can likely account for the long-term trend only.

Exciting new results have recently come thanks to the {\it IXPE} satellite, which can measure polarization in X-rays.  
In June 2022, {\it IXPE} observed a rotation of the X-ray polarization angle of the source Mkn~421 (Di Gesu et al. 2023),  with roughly constant rotation rate, while the EVPA at lower frequencies remained constant. This suggested that the X-ray emission comes from a jet region that is distinct from that producing the low-energy radiation, probably an inner spine. There, a localized shock propagating along the helical magnetic structure of the jet caused the rotation of the polarization angle.\\

\vskip-3mm

\section{PERIODICITY}

As mentioned in the Introduction, blazars in general show unpredictable variability, but sometimes periodic patterns have been reported.
The most famous example is the BL Lac object OJ~287, which already long ago (Sillanp\"a\"a et al.~1988) was recognised to exhibit a periodicity of about 12~yr in its optical light curve. This was interpreted as due to a SMBH binary system. Actually, it was later found that the outbursts are not strictly periodic, and that they are double-peaked, with a separation of about 1-2 years between the peaks.

There are essentially two types of models that can explain the source periodicity. In the first type, the outbursts are due to true luminosity changes that are triggered by the impact of the secondary black hole with the accretion disc of the primary black hole. According to Valtonen et al. (2023), this impact occurs twice every orbit, so that both peaks of an outburst are thermal, and this means that they are visible in the optical band, but not in the radio or in X rays.
In contrast, in the model by Valtaoja et al. (2000), the first peak is due to the impact and it is thermal, while the second peak is produced by something happening inside the jet, triggered by the previous impact. Therefore, the second peak is predicted to be visible also in the radio band. One may think that it should be easy to discriminate between the two models, just by looking at radio-optical correlations. However, the comparison between the optical and radio light curves is somehow tricky, and does not lead to conclusive results.

But there is a second type of models, according to which the outbursts are of geometric nature, due to orientation changes of the jet that lead to an increase of the Doppler beaming.
Britzen et al. (2018) proposed that the radio and optical light curves are the result of jet precession, rotation and nutation, with a period that is twice that previously claimed.
The Villata et al. (1998) model assumes that both the SMBHs of a binary system have a jet, which is bent and precessing, and periodically aligns with the line of sight, causing the observed outbursts.

Another example of periodic behaviour with a shorter time scale was recognized in the optical light curves of the BL Lac object S4~0954+65 during a one year long WEBT campaign including observations by the {\it TESS} satellite (Raiteri et al. 2021b).
We found that the main peaks and dips repeat every about one month, and we suggested that they were due to a rotating helical jet. The modulation of the flare amplitude was explained in terms of a variable helix pitch angle, because of small oscillations of the helical structure around its dynamical equilibrium configuration.

\begin{figure}
	\centering
\includegraphics[width=6.2cm,keepaspectratio=true]{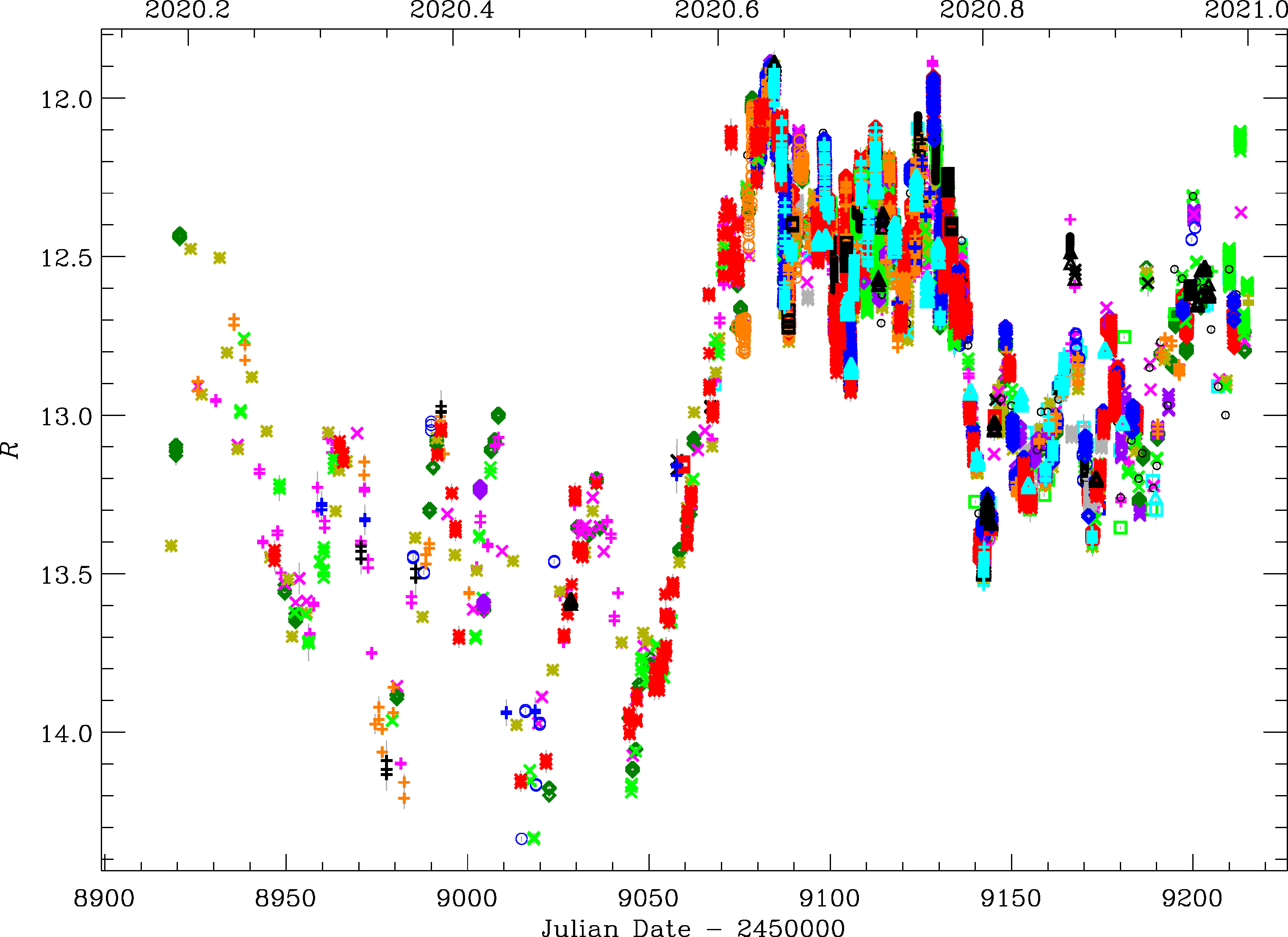}
\includegraphics[width=6.6cm,keepaspectratio=true]{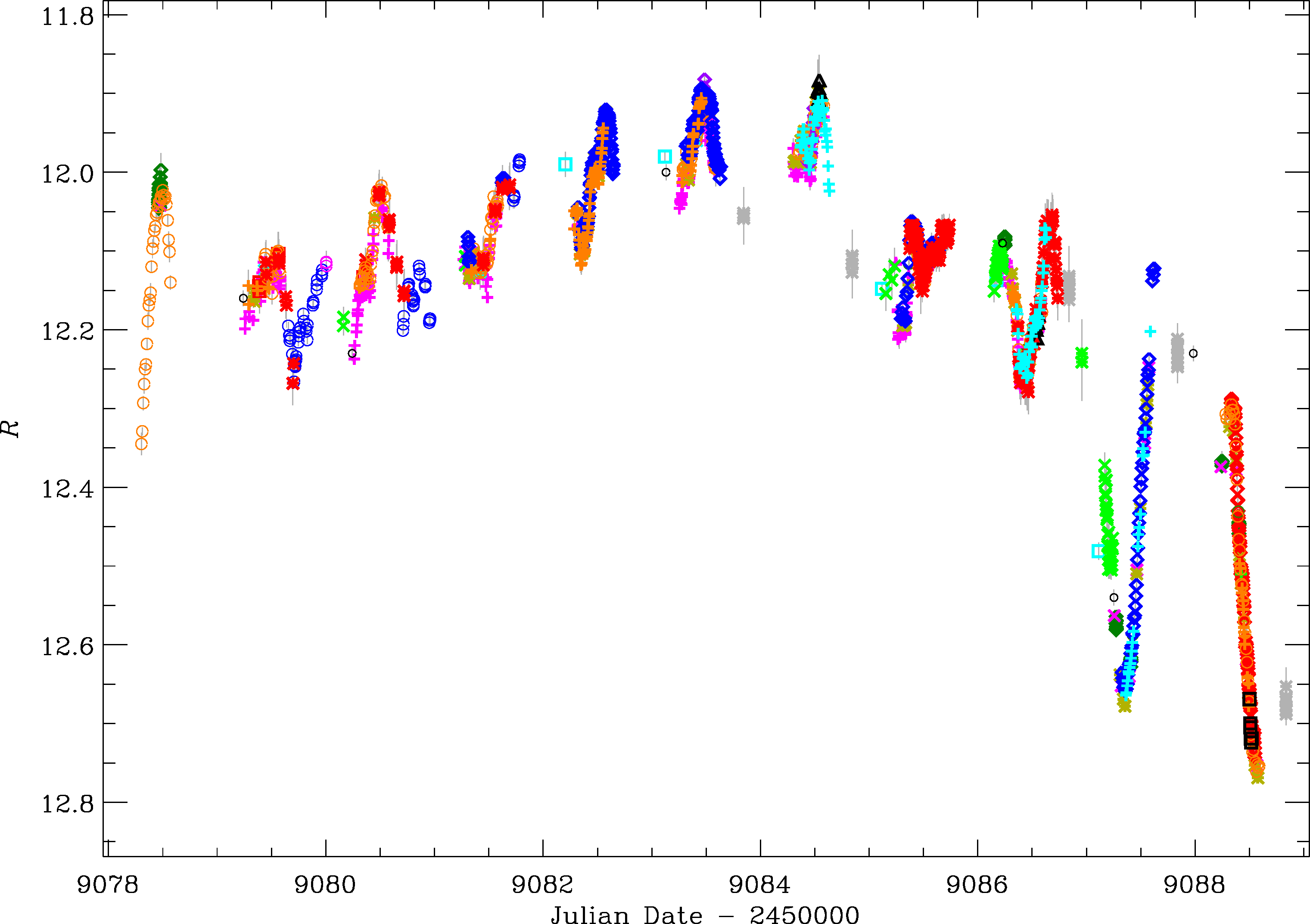}

\vspace{-3mm}

\caption{\emph{Left}: $R$-band light curve of BL Lacertae in 2020 obtained with data from the WEBT published in Jorstad et al. (2022). Different symbols and colours distinguish different datasets. \emph{Right}: An enlargement of the light curve during the first phase of the 2020 outburst, showing the detected QPOs.}
	\label{bl2020}
\vspace{-1mm}
\end{figure}

But periodicities are also found on even shorter time scales, and can be transient. We discovered rapid quasi-periodic oscillations (QPOs) in the relativistic jet of BL Lacertae during a WEBT campaign in 2020 (Jorstad et al. 2022). In the first phase of the outburst, a transient quasi-periodicity of about 13 hours was identified thanks to the exceptional sampling (see Figure \ref{bl2020}). This periodicity was also seen in the $\gamma$-ray light curve and in the polarization behaviour. We combined the optical and $\gamma$-ray data  with VLBA radio images at several epochs during the event. These radio images show a core, three standing shocks, some kinks in the jet structure, and a feature that was seen to propagate down the jet at relativistic speed. 
We suggested that when this feature, which is likely a moving shock, met the second standing shock, it triggered a kink instability. The kink instability, in turn, triggered the QPOs that we observed.
\begin{figure}
	\centering
\includegraphics[width=10cm,keepaspectratio=true]{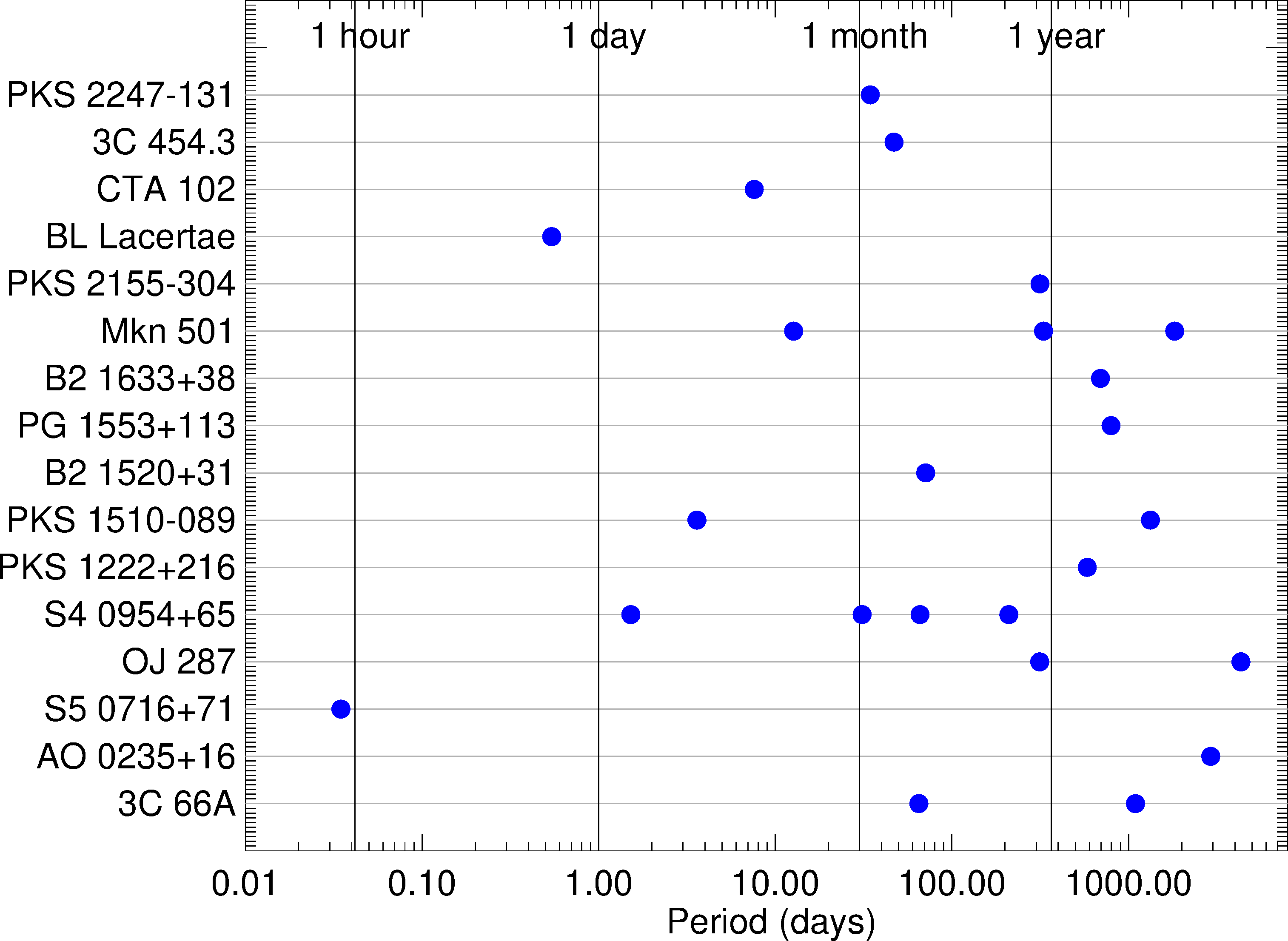}

\vspace{-3mm}

\caption{Time scales of claimed periodicities in the multiwavelength light curves of some blazars, which are identified on the left.}
	\label{period}
\vspace{-1mm}
\end{figure}

The discovery of periodic behaviour in blazars is a very important clue that can reveal what happens in the jet or at the base of the jet, in its central engine.
The literature is full of papers reporting the discovery of periodicities in blazar light curves at different frequencies: mostly in the optical, but also in the radio and $\gamma$-ray bands.
Some of these published claims are reported in Figure~\ref{period}, where the corresponding source is indicated.
Periods range from less than one hour to several years, and sometimes multiple periods are found in the same source.
Different mechanisms are proposed to explain these periodicities, involving either the jet, or the accretion disc, or the central engine.
However, as Vaughan et al. (2016) pointed out, stochastic processes can produce false periodicities, so we must be very careful in assessing the significance of these signals.
Moreover, Covino et al. (2019) analysed the $\gamma$-ray light curves of some blazars to see whether they could confirm the periodicities claimed in the literature, but they could not find any strong case, emphasizing the need to be very cautious.

\references

Abdo, A.~A., Ackermann, M., Ajello, M., et al.: 2010, \journal{Nature}, \vol{463}, 919. 

Acharya, S., Borse, N.~S., Vaidya, B.: 2021, \journal{Monthly Notices of the Royal Astronomical Society}, \vol{506}, 1862.

Aharonian, F., Akhperjanian, A.~G., Bazer-Bachi, A.~R., et al: 2007,\journal{The Astrophysical Journal}, \vol{664}, L71.

Albert, J., Aliu, E., Anderhub, H., et al.: 2007, \journal{The Astrophysical Journal}, \vol{669}, 862.

Blinov, D., Pavlidou, V., Papadakis, I, et al.: 2018, \journal{Monthly Notices of the Royal Astronomical Society}, \vol{474}, 1296.

Britzen, S., Fendt, C., Witzel, G., et al.: 2018, \journal{Monthly Notices of the Royal Astronomical Society}, \vol{478}, 3199.

Covino, S., Sandrinelli, A., Treves, A.: 2019, \journal{Monthly Notices of the Royal Astronomical Society}, \vol{482}, 1270.

D'Ammando, F., Raiteri, C.~M., Villata, M., et al.: 2019, \journal{Monthly Notices of the Royal Astronomical Society}, \vol{490}, 5300.

de Jaeger, T., Shappee, B.~J., Kochanek, C.~S., et al.: 2023, \journal{Monthly Notices of the Royal Astronomical Society}, \vol{519}, 6349.

Di Gesu, L., Marshall, H.~L., Ehlert, S.~R., et al.: 2023, \journal{Nature Astronomy}, \vol{7}, 1245.

%Fuentes, A., G\'omez, J.~L., Mart\'i, J.~M., et al.: 2023, \journal{Nature Astronomy}, https://doi.org/10.1038/s41550-023-02105-7.

Ghisellini, G., Foschini, L., Tavecchio, F., Pian, E.: 2007, \journal{Monthly Notices of the Royal Astronomical Society}, \vol{382}, L82.

Giommi, P., Glauch, T., Padovani, P., et al.: 2020, \journal{Monthly Notices of the Royal Astronomical Society}, \vol{497}, 865.

Jorstad, S.~G., Marscher, A.~P., Raiteri, C.~M., et al.: 2022, \journal{Nature}, \vol{609}, 265.

Ledden, J.~E., Aller, H.~D.: 1979, \journal{Astrophysical Journal}, \vol{229}, L1.

Lico, R., Liu, J., Giroletti, M., et al.: 2020, \journal{Astronomy \& Astrophysiscs}, \vol{634}, A87.

Marscher, A.~P., Gear, W.~K.: 1985, \journal{The Astrophysical Journal}, \vol{298}, 114.

Marscher, A.: 2014, \journal{The Astrophysical Journal}, \vol{780}, 87.

Moll, R., Spruit, H.~C., Obergaulinger, M: 2008, \journal{Astronomy \& Astrophysics}, \vol{492}, 621.

Nakamura, M., Uchida, Y., Hirose, S.: 2001, \journal{New Astronomy}, \vol{6}, 61.

Raiteri, C.~M., Villata, M., Acosta-Pulido, J.~A., et al.: 2017, \journal{Nature}, \vol{552}, 374.

Raiteri, C.~M., Villata, M., Carosati, D., et al.: 2021a, \journal{Monthly Notices of the Royal Astronomical Society}, \vol{501}, 1100.

Raiteri, C.~M., Villata, M., Larionov, V.~M., et al.: 2021b, \journal{Monthly Notices of the Royal Astronomical Society}, \vol{504}, 5629.

Raiteri, C.~M., Villata, M., Carnerero, M.~I., et al.: 2023, \journal{Monthly Notices of the Royal Astronomical Society}, \vol{526}, 4502.

Shukla, A., Mannheim, K.: 2020, \journal{Nature Communications}, \vol{11}, 4176.

Sillanp\"a\"a, A., Haarala, S., Valtonen, M.~J., et al.: 1988, \journal{Astrophysical Journal}, \vol{325}, 628.

Sironi, L., Spitkovsky, A.: 2014, \journal{The Astrophysical Journal Letters}, \vol{783}, L21.

Urry, C.~M., Padovani, P.: 1995, \journal{Publications of the Astronomical Society of the Pacific}, \vol{107}, 803.

Valtaoja, E., Ter\"asranta, H., Tornikoski, M., et al.: 2000, \journal{The Astrophysical Journal}, \vol{531}, 744.

Valtonen, M.~J., Zola, S., Gopakumar, A., et al.: 2023, \journal{Monthly Notices of the Royal Astronomical Society}, \vol{521}, 6143.

Vaughan, S., Uttley, P., Markowitz, A.~G., et al.: 2016, \journal{Monthly Notices of the Royal Astronomical Society}, \vol{461}, 3145.

Villata, M., Raiteri, C.~M., Sillanp\"a\"a, A., Takalo, L.~O.: 1998, \journal{Monthly Notices of the Royal Astronomical Society}, \vol{293}, L13. 

%Villata, M., Raiteri, C.~M., Aller, M.~F., et al.: 2007, \journal{Astronomy \& Astrophysics}, \vol{464}, L5.

Villata, M., Raiteri, C.~M., Gurwell, M.~A., et al.: 2009, \journal{Astronomy \& Astrophysics}, \vol{504}, L9.

Wagner, S.~J., Witzel, A.: 1995, \journal{Annual Review of Astronomy and Astrophysics}, \vol{33}, 163.

Weaver, Z.~R., Williamson, K.~E., Jorstad, S.~G., et al.: 2020, \journal{The Astrophysical Journal}, \vol{900}, 137.

Wehrle, A.~E., Carini, M., Wiita, P.~J., et al.: 2023, \journal{The Astrophysical Journal}, \vol{951}, 58.

Zhao, G.-Y., G\'omez, J.~L., Fuentes, A., et al.: 2022, \journal{The Astrophysical Journal}, \vol{932}, 72.

\endreferences

\end{document}